\documentstyle[prb,multicol,aps]{revtex}
\input epsf.sty

\begin{document}
\title{Polariton Local States in Periodic Bragg Multiple Quantum Well Structures}
\author{Lev I. Deych\dag, A. Yamilov\ddag, A.A. Lisyansky\ddag}
\address{{\dag}Department of Physics, Seton Hall University, South Orange, NJ 07079\\
{\ddag}Department of Physics, Queens College of CUNY, Flushing, NY}
\date{\today}
\maketitle

\begin{abstract}
We analytically study optical properties of several types of defects in Bragg multiple quantum well structures. We show that a single defect leads to two local polariton modes in the photonic band gap. These modes lead to peculiarities in reflection and transmission spectra. Detailed recommendations for experimental observation of the studied effects are given.
\end{abstract}

\begin{multicols}{2}

\noindent
It has been demonstrated recently \cite{Khitrova} that long multiple quantum well (MQW) systems can form optical lattices, in which different quantum wells (QWs) are coherently coupled due to interaction with a retarded electromagnetic field. Light-matter interaction in such systems depends upon their structure and can be significantly and controllably modified.  Polariton formalism provides an adequate self-consistent way to describe strong interaction of the QW excitons and the light in MQW systems.\cite{Citrin,Andreani} These systems have become a subject of very active research in the past few years (see, for instance, Refs.\onlinecite{Citrin,Andreani,Stroucken} and references therein). Special attention has been paid to so called Bragg structures, where  the interwell spacing, $a$, is exactly equal to the half-wavelength of light at the frequency of excitonic resonance,   $\lambda _{0}/2=a$. \cite {Khitrova,Hubner,Vladimirova}  Pecularities of the Bragg structures follow from the fact that a photonic band gap in the vicinity of the exciton frequency is degenerate. In other words, it is formed by two adjacent gaps with coinciding boundaries.  Detuning the structure from the exact Bragg condition shifts those boundaries away from each other, giving rise to a conduction band between them.\cite{DeychQW}

Should the periodicity in the arrangement of MQWs be locally altered, one could expect the appearence of defect local modes inside the photonic bandgaps. This phenomenon provides additional possibilities to control optical properties of MQWs, and, therefore, is of considerable interest. This idea was put forward in Ref.\onlinecite{Citrinlocal} , where a dispersion equation for frequencies of the local modes with different polarizations was derived.  In the case of T-polarized excitations, the equation describing MQWs is essentially equivalent to a model of one-dimensional chain of dipoles used in our previous works to discuss local polariton states in polar crystals. \cite{Deych,PRBlocal,JOSAB} 
In the context of MQWs, the local polariton states considered in Refs. \onlinecite{Deych,PRBlocal,JOSAB} correspond to a mode localized  in the growth direction of the MQW structure, but extended in the in-plane directions. 

In this letter we study local defect polariton states in Bragg MQW structures, and defect induced changes in transmission and reflection spectra. Defect layers can differ from the host layers in three different ways: in the exciton-light coupling strength ($\Gamma $-defect), in the exciton resonance frequency ($\Omega $-defect), and in interwell spacing ($a$-defect).  We shall show below that each of these types play distinctly different roles in the optical properties of the system. This fact justifies consideration of these three situations separately, even though it may be difficult experimentally to create a defect layer with a different exciton resonance frequency but the same coupling strength. At the same time the $a$-defect can  obviously be realised in its pure form. We obtain closed analytical expressions for respective local frequencies, as well as for reflection and transmission coefficients. On the basis of the results obtained, we give practical recommendation for experimental observation of the studied effects in samples used in Refs.\onlinecite{Khitrova,Hubner}.

Optical properties of QWs are usually described with the use of non-local susceptibility determined by energies and wave functions of a QW exciton.\cite{Citrin,Andreanireview} In the case of very thin QWs, a simplified approach is possible, in which the polarization density of the QW is presented in the form $P({\bf r},z)=P_n({\bf r})\delta(z-z_n)$, where ${\bf r}$ is an in-plane position vector,  $z_n$ represents a coordinate of the $n$th well, and $P_n$ is a surface polarization density of the respective well.  When light is incident in the direction of growth $z$ of MQWs,  $k_{||}=0$ and there exist two independent degenerate transversal polarizations, $T$ and $L$, which are not coupled to the longitudinal $Z$ mode. In this case, the dynamics of transverse modes can be descibed by  equations 
\begin{equation}
\left( \Omega _{n}^{2}-\omega ^{2}\right) P_{n}=(c/\pi )\Gamma _{n} E(z_{n}),
\label{P}
\end{equation}
\begin{equation}
\frac{\omega ^{2}}{c^{2}}E\left( z\right) +\frac{d^{2}E\left( z\right) }{ dz^{2}}=-4\pi \frac{\omega ^{2}} {c^{2}} \sum\limits_{n} P_{n} \delta \left( z-z_{n}\right),
\label{E}
\end{equation}
which coincide with equations used in Refs. \onlinecite {Deych,PRBlocal,JOSAB,Deutsch} to describe one-dimensional chains of atoms.
Here $\Omega _{n}$ and $\Gamma _{n}$ are the excitonic frequency and exciton-light coupling of the $n$th QW, respectively.  In an infinite pure system all $\Gamma _{n}=\Gamma _{0},\ \Omega _{n}=\Omega _{0}$, and $z_{n}=na=n\lambda _{0}/2$. The spectrum of ideal MQWs has been studied in many papers.\cite{Citrin,Andreani,DeychQW,Deutsch,Ivchenko2} In the specific case of Bragg structures, the exciton resonance frequency is at the center of the  bandgap determined by the inequality  $\omega _{l}=\Omega _{0}( 1-\sqrt{2\Gamma _{0} / \pi \Omega _{0}}) <\omega <\Omega_{0}( 1+\sqrt{2\Gamma _{0}/ \pi \Omega _{0}}) =\omega _{u}.$\cite{DeychQW} This bandgap is the frequency region where we will look for new local states associated with defects in MQWs.

$\Omega$- and $\Gamma$-defects introduce perturbations in the equation of motion that are localized at one site (diagonal disorder). Therefore, they can be studied by the usual Green's function technique (see, for instance, Ref.\onlinecite{Lifshitz}). The resulting dispersion equations have the form 
\begin{equation}
G_ {_{\Omega,\Gamma}}=\beta/2\sqrt{D},
\label{cond_inf_Om}
\end{equation}
where $\beta =4\Gamma _{0}\omega /\left(\omega ^{2}-\Omega _{0}^{2}\right)$ and $D=-1 + \beta ^{2}/4 + \beta \cot (\omega a/c)$. For the $\Omega$-defect the function $G_{_{\Omega}}=\left(\Omega _{1}^{2}-\omega ^{2}\right)/\left(\Omega _{1}^{2}-\Omega _{0}^{2}\right) $ and for the $\Gamma$-defect  the respective function is $G_{_{\Gamma}}=\Gamma_{0}/\left(\Gamma_{1}-\Gamma_{0}\right)$. $\Omega_1$ and $\Gamma_1$ denote respective parameters of the defect layer. A similar equation for the $\Omega$-defect has been studied in Ref. \onlinecite{Deych} in the longwave approximation. It was found that the equation has one real value solution for any $\Omega_1>\Omega_0$. In the case of Bragg structures there are always two solutions for both types of the defects, one below $\Omega_1$ and one above. This is a manifestation of the degenerate nature of the bandgap in Bragg structures. The above equations can be solved approximately using the fact that $\Gamma _{0} \ll \Omega_0$ in most cases. For the $\Omega$-defect, one solution demonstrates a radiative shift from the defect frequency $\Omega_1$ 
\begin{equation}
\omega _{def}^{(1)}=\Omega _{1}-\Gamma _{0}(\Omega _{1}-\Omega _{0})/\sqrt{\left( \omega _{u}-\Omega _{1}\right) \left( \Omega _{1}-\omega _{l}\right) },
\label{cond_inf_Om_app}
\end{equation}
while the second solution  splits off the upper or lower boundary depending upon the sign of $\Omega _{1}-\Omega_{0}$:  
\begin{equation}
\omega_{def}^{(2)}=\omega _{u,l} \pm  \pi^2 (\omega _{u}-\omega _{l})\left(\Omega _{1}-\Omega _{0} \right)^2/4\Omega _{0}^{2}, \label{def2freq}
\end{equation}
where one chooses $\omega_u$ and ``$-$" for $\Omega _{1}>\Omega_{0}$, and $\omega_l$ and ``$+$" in the opposite case.
In the case of the $\Gamma$-defect, both solutions appear in the vicinity of the gap boundaries $\omega _{def}^{(1,2)}=\omega _{u,l} \pm 2\left(\Gamma_1-\Gamma_0\right)^{2}\left(\omega _{u}-\omega _{l}\right)$.
These solutions  exist only for $0<\Gamma _{1}<\Gamma _{0}$ and are  very close to the gap boundaries. One could expect, therefore, that the states at these frequencies are vulnerable to even a weak dissipation, and would not significantly affect optical spectra of the system. 

The $a$-defect significantly differs from the two other types. An increase in the interwell distance between any two wells automatically changes the coordinates of an infinite number of wells: $z_{n}=na$ for $n\leq n_{d}$ and $z_{n}=(b-a)+na$ for $n_{d}<n$, where $b$ is the distance between the $n_{d}$th and $(n_{d}+1)$th wells. Therefore, this defect is non-local and cannot be treated using the same methods as in two previous cases. The best approach to this situation is to match solutions of semi-infinite chains for $n<n_d$ and $n>n_d+1$ with a solution for $na<z<na+(b-a)$. Solutions for semi-infinite chains can be constructed using the transfer matrix approach, in which the state of the system is described by a two-dimensional vector  $v_{n}$ with components $E(z_{n})$ and $(c/\omega)dE(z_{n})/dz$. Propagation of this vector through the system is described by the transfer matrix $\hat\tau_{n}$:
\begin{equation}
\widehat{\tau _{n}}=\left( 
\begin{array}{cc}
\cos (\frac {\omega }{c}a_{n})+\beta \sin (\frac {\omega }{c}a_{n}) & \sin (\frac {\omega }{c}a_{n}) \\ 
-\sin (\frac {\omega }{c}a_{n})+\beta \cos (\frac {\omega }{c}a_{n}) & \cos (\frac {\omega }{c}a_{n})
\end{array}
\right), 
\label{matrix}
\end{equation}
where $a_{n}=z_{n+1}-z_{n}$. As a result, one obtains the dispersion equation for the defect mode in terms of elements of the total transfer matrix $\hat{T}$, equal to the product of all site matrices $\hat\tau$:
\begin{equation}
\left( T_{11}+T_{22}\right) -i\left( T_{12}-T_{21}\right) =0.
\label{quasi_mode}
\end{equation}
In the limit of an infinitely long system, the imaginary part of this equation vanishes, and one has a real valued dispersion equation for the frequency of a stationary local mode.  Using Eq. (\ref{matrix}) one can present Eq. (\ref{quasi_mode}) for an infinite MQW system as
\begin{equation}
\cot(\frac{\omega}{c} b)=-\left[\sin (\frac{\omega}{c}a)-\beta \lambda_-/2 \right]/ \left[\cos (\frac{\omega}{c}a)-\lambda_- \right],
\label{cond_inf_a}
\end{equation}
where $\lambda _{-}=\left[ \cot (\omega a/c)+\beta/2 - \sqrt{D}\right] \sin (\omega a/c)$ is one of the eigenvalues of the transfer matrix Eq. (\ref{matrix}). This equation also has two solutions - above and below $\Omega_0$. Assuming that $\sqrt{\Gamma_0}b/\Omega_0 a \ll 1$ one of these solutions can be expressed as
\begin{equation}
\omega _{def}^{(1)}=\Omega _{0}-\frac{\omega _{u}-\omega _{l}}{2}\frac{(-1)^{ \left[ \frac{\xi +1}{2} \right] } \sin { \frac{\pi }{2}\xi  }}{1+ \frac{\omega _{u}-\omega _{l}}{2\Omega _{0}}\frac{b}{a}  (-1)^{ \left[ \frac{\xi +1}{2} \right] } \cos { \frac{\pi }{2}\xi }  },
\label{om1_a_def}
\end{equation}
where $\xi =b/a$, and $[...]$ denotes an integer part. The second solution can be obtained from Eq. (\ref{om1_a_def}) by replacing $\xi $ by $\xi+1$. Therefore, for $\Gamma_0\ll\Omega_0$ and not very large $\xi$, both solutions are almost periodic fuctions of $b/a$ with the period of $1$.

The expression on the left hand side of Eq. (\ref{quasi_mode}) coincides with the denominator of the transmission and reflection coefficients in a system of finite length, and with the appropriate choice of transfer matrices, $\tau$, equation $T_{11}+T_{22}=0$ produces dispersion equations for local states of all three types of defects. In the absence of homogeneous broadening of the exciton resonance, the defects would cause a resonance increase in transmission at the local mode frequency.\cite{Deych} The resonance occurs when the defect is placed at the center of the system. Then the maximum transmission becomes independent of the system's length, and in the case of $\Omega$- and $\Gamma$-defects it can be presented as 
\begin{equation}
|t_{\max }|^{2}=1-4\left[ \left( \frac{\omega _{def}-\Omega _{0}}{\omega _{u}-\Omega _{0}}\right) ^{2}-\frac{1}{2}\right]^2.
\label{tmax}
\end{equation}
In the absence of absorption, transmission reaches unity if the frequency of the local state is $\omega _{def}=\Omega _{0}\pm \left( \omega _{u}-\Omega _{0}\right)/\sqrt{2}$. This cannot happen for the $\Gamma $-defect, because frequencies of the respective local states always lie close to the band edge, but for the $\Omega $-defect it is quite possible to create the state with the required frequency.

For the third type of defect, the resonance transmission also takes place when the defect is in the center of the stack. $t_{\max }$, in this case, can be expressed as 
\begin{equation}
|t_{\max }|^{2}=8\left[ \frac{\omega _{def}-\Omega _{0}}{\omega _{u}-\Omega _{0}}\left( 1-\left( \frac{\omega _{def}-\Omega _{0}}{\omega _{u}-\Omega _{0}}\right) ^{2}\right) \right]^2.
\label{tmax_a_def}
\end{equation}
It becomes unity for two symmetric with respect to the center of the gap frequencies: $\omega _{def}^{(1,2)}=\Omega _{0}\pm \left( \omega _{u}-\Omega _{0}\right)/\sqrt{2} $. As one can see from Eq. (\ref{om1_a_def}), these conditions can be satisfied for both defect frequencies at the same time when $b \simeq (integer+1/2)a$. 
\begin{figure}
\epsfxsize=3.2in \epsfbox{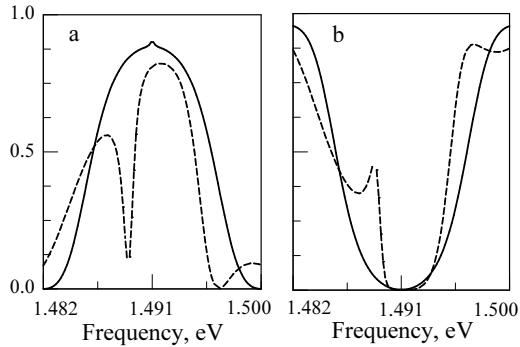}
\vspace{-2.1in}
\caption{Reflection (a) and transmission (b) coefficients of the 200 QWs with defect placed in the middle of the stack. Solid lines correspond to the pure system, and dashed lines correspond to the system with the $a$-defect ($b/a=1.3$).}
\end{figure}
In a real system, enhancement of the transmission coefficient is usually limited by homogeneous broadening. Two cases are possible when exciton damping is taken into account. It can suppress the resonance transmission, and the presence of the local states will only be observed as an enhancement of absorption at the local frequency. This can be called a weak coupling regime for the local state, when incident radiation is resonantly absorbed by a local exciton state. The opposite case, when the resonance transmission persists in the presence of damping, can be called a strong coupling regime. In this case, there is a coherent coupling between excitons and the electromagnetic field, so that the local state can be suitably called a local polariton. Among three considered types of defect, the $\Gamma$-defect is less likely to survive absorption because of the proximity of the respective frequencies to bandgap edges. For the $\Omega$-defect one of the local frequencies appears far enough from the boundaries, and can be less sensitive to absorption. However, the width of the respective transmission resonance is determined by its radiative shift from $\Omega_1$, where transmission goes to zero. This shift is rather small and small absorption can still suppress the resonance transmission. Therefore, the best candidate to produce a local polariton state in the strong coupling regime is the $a$-defect. 

To account for homogeneous broadening quantitatevly, we add an imaginary part to the exciton polarizability, $\beta =4\Gamma _{0}\omega/\left(\omega^{2}-\Omega _{0}^{2}+2i\gamma \omega\right)$. For numerical calculations we use parameters from Ref. \onlinecite{Khitrova}. The localization length at the center of the forbidden band gap is in this case $\sim 80\cdot a$, while the length of the samples used reached $100\cdot a$. Fig. 1 presents plots of reflection and transmission for a MQW system with an $a$-defect, for which the resonance transmission is the most pronounced. We can conclude that the interwell spacing defect gives rise to local polariton states in regular MQW $InGaAs/GaAs$ MQWs. These states manifest themselves in strong resonant tunneling of light through a MQW system with $100$ or more wells and can be observed in transmission experiments. This type of defect can be implemented experimentally and present additional opportunities for controlling light-matter interaction with a potential for practical applications.

We are indebted to S. Schwarz for reading and commenting on the manuscript. This work was partially supported by NATO 
Linkage Grant N974573 and PSC-CUNY Research Award.

\end{multicols}

\end{document}